\documentclass[aps,11pt]{revtex4-2}

\usepackage{graphicx}

\usepackage{float}
\usepackage[caption=false]{subfig}

\usepackage{amsmath}

\usepackage{upgreek}

\newcommand{\mr}{\mathbf{r}}

\usepackage{tabularx}
\usepackage{dcolumn}

\usepackage{upgreek}

\newcommand{\mft}{\tilde{\mathbf{f}}}
\newcommand{\mgt}{\tilde{\mathbf{g}}}
\newcommand{\tlo}{\tilde{\omega}}
\newcommand{\ud}{\mathrm{d}}

\begin{document}

\title{\Large{A lithographically-defined quantum dot with sub-wavelength confinement of light}}

\author{George Kountouris $^{1,2}$, Anne Sofie Darket $^{1,2}$, Lea Vestergaard $^{1,2}$, Emil Vosmar Denning$^{1,2,3}$, Jesper M{\o}rk$^{1,2}$, and Philip Tr{\o}st Kristensen$^{1,2}$ \\
\vspace{5mm}
\scriptsize{1. Department of Electrical and Photonics Engineering, Technical University  of  Denmark, DK-2800  Kgs.   Lyngby,  Denmark} \\
\vspace{3mm}
\scriptsize{2. NanoPhoton -- Center for Nanophotonics, Technical University  of  Denmark, DK-2800  Kgs.   Lyngby,  Denmark} \\
\scriptsize{3. Institut f{\"u}r Theoretische Physik, Technische Universita{\"a}t Berlin, 10623 Berlin, Germany}}

\begin{abstract}
We present an optical cavity with deep sub-wavelength confinement of light in a region that simultaneously works as a quantum dot. The design is based on a dielectric membrane with a buried quantum well and restricts the electron and hole wave functions to the area of the optical hotspot in order to overcome the challenge of co-locating an optical cavity with a quantum emitter. Combined with proper surface passivation, this geometry points towards the deterministic fabrication of functional quantum dots in optical cavities by lithographic means.
\end{abstract}


\maketitle

\vspace{-15mm}

\section{Introduction}

Semiconductor quantum dots (QDs) have enabled on-demand single-photon emission with high brightness and high purity \cite{Claudon2010, Ding2016, Schweickert2018}. This combination of features makes them a very promising quantum light-matter interface for quantum key-distribution~\cite{BassoBasset2021, Bozzio2022, Vajner2022}, quantum networks~\cite{Lodahl2017, Zhu2020, Vajner2022}, or quantum computing~\cite{Gao2012, Coste2023}, but deterministic fabrication of single QDs with specified transition energies and well-defined positions is an important challenge to be overcome for practical future applications. Current successful approaches typically rely on either alignment-based methods \cite{Dousse2008, Schneider2012, Jns2012, Sapienza2015, Coles2016, Calic2017, Strau2017, Sartison2017, Liu_2017} or transfer techniques \cite{Junno1995, Toset2006, vanderSar2009, AmpemLassen2009, Reimer2012, Zadeh2016, Kim2017, Davanco2017, Katsumi2018}. In the former approach, QDs with desired emission properties are located and a photonic structure is subsequently fabricated around them. In the latter, the located QDs are transferred to the desired region, possibly for further fabrication steps. Recent in-situ techniques improve alignment accuracy by performing the fabrication and preselection of QDs and the fabrication of the optical structure in the same machine \cite{Dousse2008, Gschrey2013}. These approaches, however, still involve emitters grown using self-organizing techniques, which inevitably lead to random positions and a distribution of sizes and thereby impact the scalability. As an alternative to alignment-based methods or transfer techniques, lithographical patterning and etching of quantum well-structures is a well-known method of fabricating QDs with nanometer-scale control of both size and position. However, the method has historically resulted in QDs with prohibitively high non-radiative rates, which
greatly suppress the emission and degrade performance~\cite{Clausen1989}, suggesting that such top-down approaches are very difficult for small QDs. Nevertheless, Hiriyama \emph{et al.}~\cite{Hirayama1994} were able to achieve lasing with 30nm-diameter lithographically-defined QDs (LDQDs) and Verma \emph{et al.}~\cite{Verma_OSA_2011} have successfully demonstrated luminescence with LDQDs of similar size. Compared to self-assembled QDs, LDQDs still suffer from a relatively large linewidth and inferior performance when considering the single-photon emission properties as gauged by the antibunching performance. Both shortcomings can be attributed to non-radiative recombination at the etched surfaces. Surface treatment and passivation may be one way to improve performance \cite{Clausen1989, Andrade2021}, but -- to the best of our knowledge -- this has so far been unsuccessful in creating QDs with narrow linewidths and antibunching characteristics competitive with those of self-assembled QDs.

\medskip

In this work, we investigate the use of dielectric optical cavities with extreme confinement of light \cite{Gondarenko2006, Hu2016, Choi2017, Hu2018, Wang2018, Albrechtsen2022ncoms} to overcome the limitations of LDQDs due to non-radiative decay. Placing QDs in photonic nanostructures is a well-established technological method for improving performance by guiding emission \cite{Ding2016, Daveau2017} and enhancing the light-matter interaction via an increased optical field strength. For a point dipole emitter, the effect is quantified by the relative enhancement of the radiative emission rate compared to an unstructured medium. In optical cavities with effective resonance wavelength $\lambda/n$, in which $n$ is the refractive index, this ratio can be calculated by the Purcell formula \cite{Purcell1946},
\begin{equation}
\label{eq1}
F_{\rm{p}} = \frac{3}{4\pi^2} \left( \frac{\lambda}{n}\right)^3 \frac{Q}{V_{\text{eff}}},
\end{equation}
where $Q$ is the quality factor, and $V_\text{eff}$ is the effective mode volume~\cite{Kristensen2012}. This expression is valid in the bad cavity limit, where the cavity decay rate is the largest dissipation rate of the system \cite{Mork2018}. Equation (\ref{eq1}) succinctly shows that improving the temporal and spatial confinement of the field will lead to an increase in the radiative enhancement. In the presence of non-radiative rates, therefore, Purcell enhancement can be leveraged to make the radiative process dominate and thereby improve the efficiency and indistinguishability of emitted photons~\cite{IlesSmith2017}.

\medskip

Historically, there have been two main paradigms for electromagnetic resonators. Plasmonic structures enable extreme spatial confinement of the electromagnetic field but suffer from high non-radiative rates, which leads to irreversible loss of energy as well as heating \cite{Wang2006, Govorov2007, Boriskina2017}. Dielectric cavities, on the other hand, offer large quality factors but have traditionally offered significantly larger effective mode volumes. Dielectric bowtie cavities, however, have emerged as a new class of devices capable of pushing the effective mode volume deep below the cubic half wavelength $(\lambda/2n)^3$ and can therefore reach very high Purcell enhancement rates even with relatively low quality factors, making them appealing for broadband applications \cite{Mork2020}. This combination of high Purcell enhancement with low $Q$ is particularly interesting for QD applications since it facilitates good spectral overlap while maintaining high Purcell rates. Placing a stochastically grown quantum emitter in the nanometer-scale hotspot of a dielectric bowtie cavity remains an enormously challenging task, however. We propose, therefore, the lithographic definition of a single QD in a dielectric bowtie cavity, and we provide a design of a bowtie cavity such that it forms an LDQD in which the exciton states are co-localized with the hotspot of the optical cavity mode. By quantitative analysis of the light-matter interaction, we find that this deterministic fabrication approach is a promising way to leverage the high radiative enhancement in dielectric bowtie cavities and thereby overcome the non-radiative decay in etched QDs.

The article is organized as follows: First, we describe the design for an etched QD in a dielectric bowtie cavity and the general design principles. Then, we introduce the modeling framework for the optical field and the electronic states and calculate the coupling strength in the limit of a single exciton interacting with a single optical cavity mode. Using this coupling strength, we then quantify the radiative rates and efficiency of designs operating in the bad cavity limit. Finally, we present our conclusions.

\section{LDQD Design}
\label{sec_LDQD_design}

\begin{figure}
  \centering
    \subfloat[\label{fig1a}]{\includegraphics[width=0.37\columnwidth]{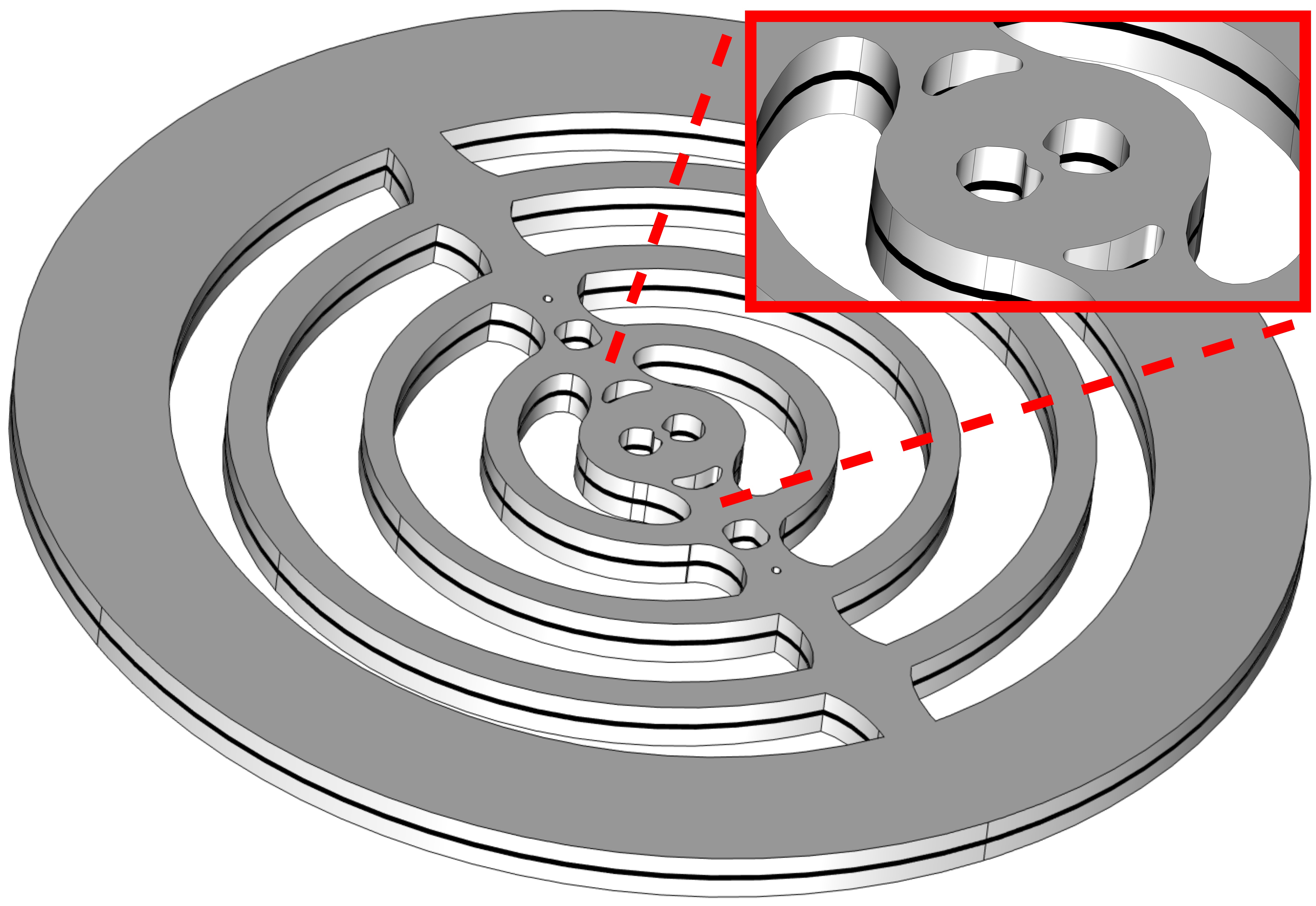}} \;\;\;
    \subfloat[\label{fig1b}]{\includegraphics[width=0.58\columnwidth]{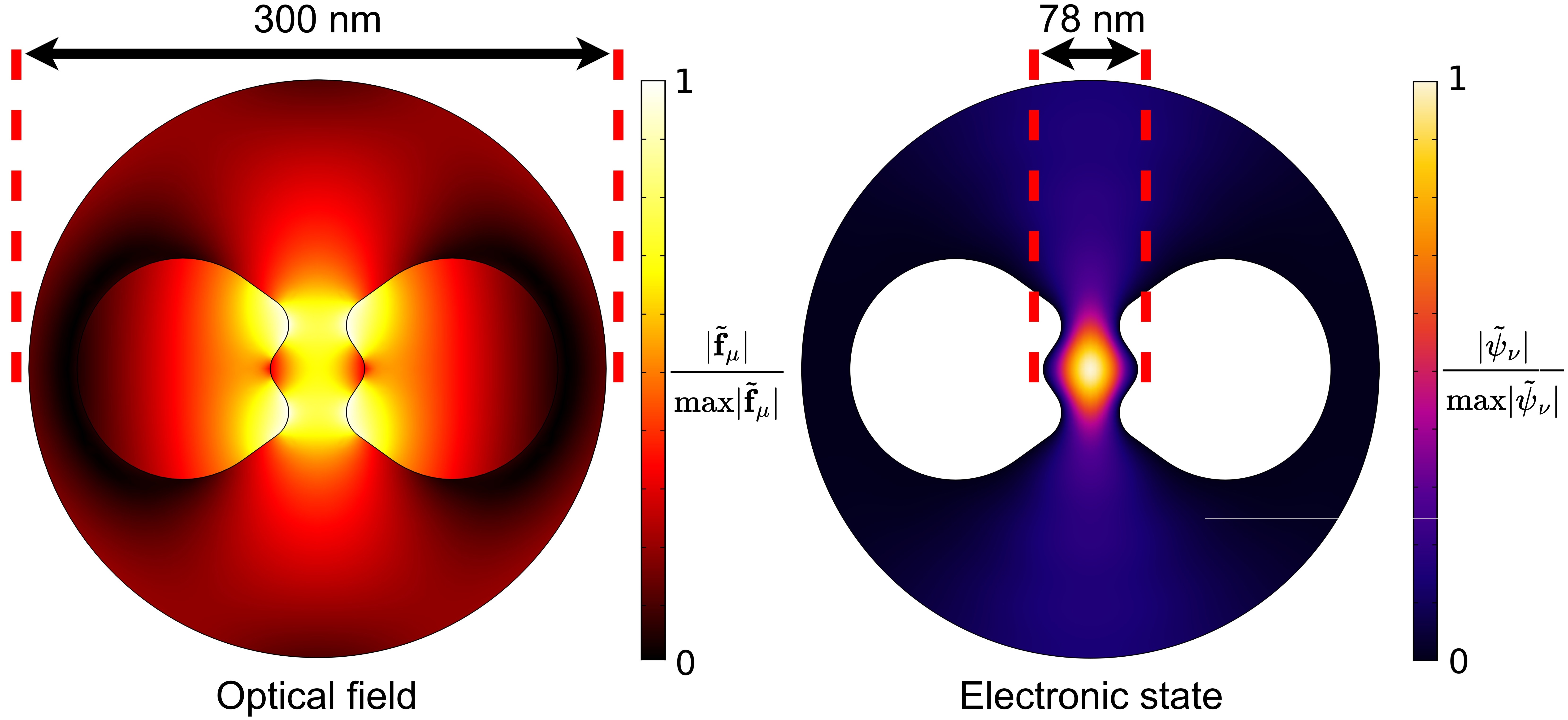}}
  \caption{(a): An overview of the design for realizing simultaneous deep sub-wavelength confinement of light and electronic states. A membrane with an embedded quantum well is etched in a two-dimensional pattern to realize the LDQD in the central part, as shown in the inset. (b): The LDQD formed by modifying the bowtie bridge in order to accommodate the optical field and the finite-lifetime electron and hole states. The figures show the relative norm of the electric field and the electronic state in a radius of 300nm around the center of the structure.}
  \label{fig1}
\end{figure}

Dielectric bowtie cavities achieve strong local fields by exploiting the electromagnetic interface conditions in a combination of geometry -- the bowtie shape -- and high index contrasts, as discussed in several works \cite{Almeida2004, Robinson2005, Hu2016, Choi2017, Albrechtsen2022}. In a typical design, the bowtie features are on the scale of nanometers, with the field dominantly confined in that same nanoscopic region. The localized nature of the hotspot coupled with these fine features means that introducing a single quantum emitter to the cavity is an ongoing challenge. Beyond carefully choosing a material platform and developing better fabrication methods, a proper design will play an important part in successful implementations. One prominent design approach is topology optimization \cite{Liang2013, Jensen2010, Christiansen2021a, Christiansen2021b}, which can be used to find locally optimized structures in a constrained domain by maximizing specific figures of merit. This method has proven extremely successful, but the process can be numerically demanding and time-consuming. In a previous work \cite{Kountouris_22}, we have shown that it is possible to modify the topology-optimized photonic cavity design of Ref. \cite{Albrechtsen2022ncoms} in order to obtain a simplified design with very similar characteristics. This simplification strategy is, therefore, an efficient way to design novel devices for different applications. In this way -- but now starting from a design in indium phosphide~\cite{Wang2018} -- we modify the corresponding simplified design with an embedded quantum well and redesign the bowtie to form an LDQD in the region of the photonic hotspot.

\smallskip

The LDQD is formed by stretching the central part of the bowtie bridge, as shown in Fig. \ref{fig1}.  The modification effectively forms a potential barrier to electronic states trapped in the central region, that would otherwise move freely in the quantum well. As the central region becomes more isolated, the tunneling rate of the state decreases, resulting in a longer lifetime.   This modification thus leads to the formation of lossy resonant electron and hole states that, when excited, will form an exciton through Coulomb interaction.

\smallskip

\begin{figure}
  \centering
    \subfloat{\includegraphics[width=0.7\columnwidth]{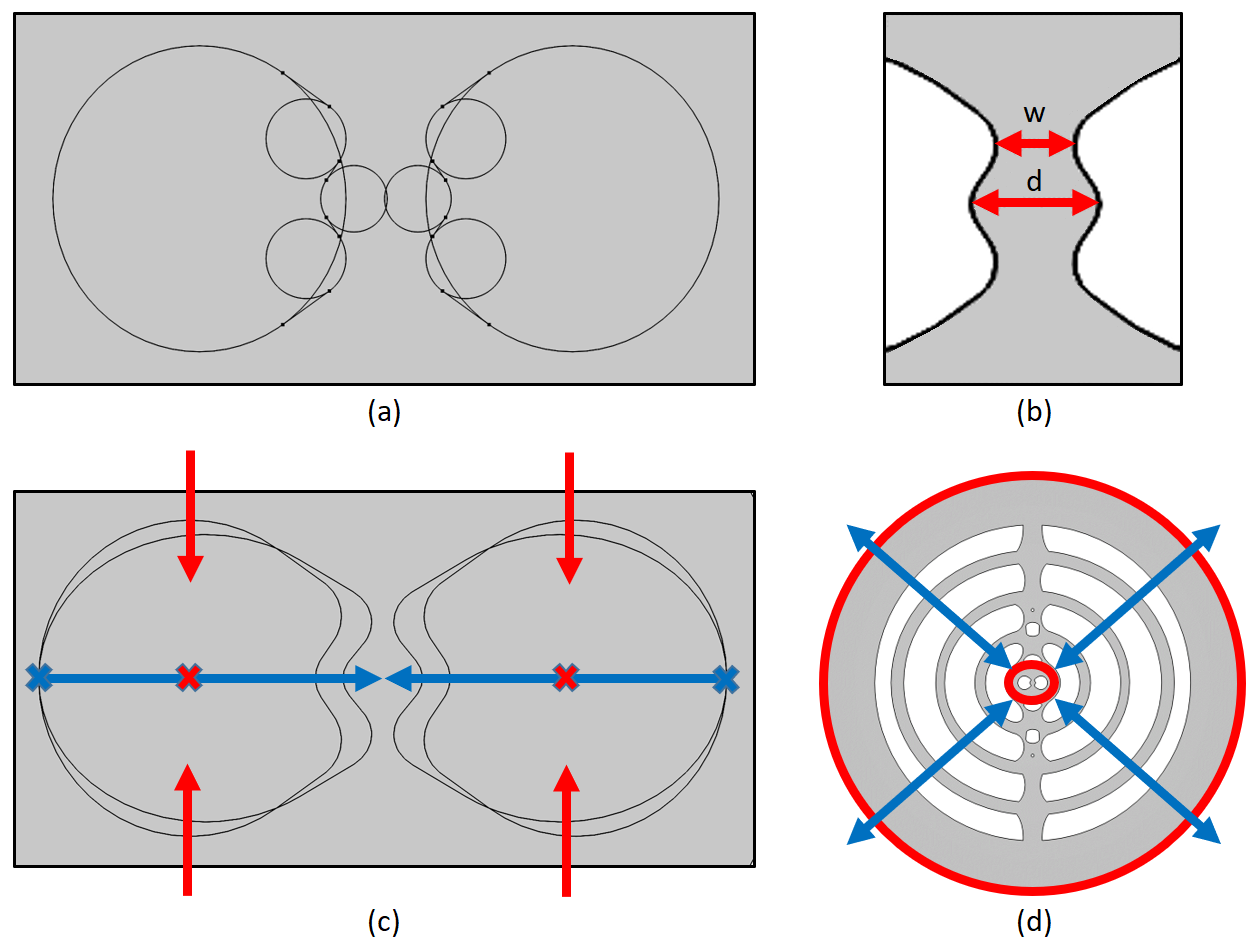}}
  \caption{Illustration of the design approach. (a). The design as defined by a set of ellipses and tangents. (b). The QD width $d$ and the bridge width $w$. (c). Scaling of the design to vary the ratio  $w/d$ for the analysis in section~\ref{sec3_4}, in which the ``x'' marks the center of scaling on each axis. (d). Scaling the design beyond the central QD region to ensure spectral overlap between the optical field and electronic states}
  \label{fig2}
\end{figure}

Modifying the bowtie shape perturbs the optical field pattern as well as the resonance frequency. To achieve strong light-matter coupling, both the spatial and the spectral overlap between the optical field and the electronic states thus needs to be considered. While formal optimization of the design may be possible using methods such as topology optimization, here we show that creating a design with a good spatial and spectral overlap is possible by using basic intuitive design choices. During our practical design process, we try to address several important considerations. First, to keep the designs fabricable, we define the features using ellipses, with curvatures that should conform to the fabrication constraints dictated by the chosen platform, as illustrated in Fig. \ref{fig2} (a). After defining the geometry, the resonances and loss rates of the electronic states can be controlled by changing the two characteristic sizes of the design, which we denote the QD width $d$ and the bridge width $w$, as shown in Fig. \ref{fig2} (b). Doing so, however, will also lead to relatively large shifts in the resonance of the optical cavity mode. In order to tune the optical resonance while minimally affecting the electronic state, we perform an overall isotropic scaling in the xy-plane of the rest of the design, as illustrated in panels (c) and (d) of Fig. \ref{fig2}. In total, the resulting geometry will define the spatial and spectral overlap, the cavity loss rate, and the QD tunneling rates. At the same time, the fabrication approach and platform are expected to contribute with additional non-radiative decay and dephasing rates. Whereas the non-radiative decay and dephasing rates are not easy to address numerically or analytically, both the tunneling lifetimes and the spatial overlap can be naturally described, for arbitrary geometries, within the chosen modeling framework as discussed below.

\section{Modeling using quasinormal modes}

We model both the electromagnetic and electronic problems in terms of quasinormal modes (QNMs) \cite{Ching1998, Kristensen2013, Lalanne2018, Kristensen2020}, also known as resonant states \cite{Muljarov2010, Both2021}. In the photonics literature, QNMs have been successfully used to describe the resonances and response of lossy optical resonators. However, the concept of QNMs is far more general and historically predates QNM applications in optics. In particular, the resonant electronic states defined from the Schr\"{o}dinger equation can also be described by QNMs \cite{GarcaCaldern1976, GarcaCaldern2010}.

The QNMs are defined as solutions to the sourceless wave equation subject to suitable radiation conditions \cite{Kristensen2020}. For the optical fields, we consider the electric field QNMs $\mathbf{\tilde{f}_\mu(r)}$, with discrete index $\mu$, and solve the corresponding wave equation along with the Silver-M{\"u}ller radiation condition,
\begin{equation}
\nabla\times\nabla\times \mathbf{\tilde{f}_\mu(r)} - \tilde{k}_\mu^2\epsilon_\text{r}(\mathbf{r})\mathbf{\tilde{f}_\mu(r)} = 0, \;\;\;\;\;\; \hat{\mathbf{r}}\times\nabla\times \mathbf{\tilde{f}_\mu(r)}\rightarrow -{\rm{i}} n_{\rm{B}} \tilde{k}_\mu\mathbf{\tilde{f}_\mu(r)}, \; |\mathbf{r}|\rightarrow \infty,
\end{equation}
\noindent where $\tilde{k}_\mu = \tilde{{\omega}}_\mu/\text{c}$ is the wavenumber, and $n_{\rm{B}}$ is the refractive index of the surrounding medium. The QNMs have complex eigenfrequencies $\tilde{{\omega}}_\mu = \omega_\mu - \text{i}\gamma_\mu$, in which the imaginary part corresponds to a characteristic decay rate.  We note that the wavenumber is complex, which results in spatially divergent field profiles in the directions away from the resonator. Nevertheless, the QNMs can be normalized using one of several complementary formulations~\cite{Kristensen2015}, see Appendix~\ref{AppendixA} for details.

\begin{figure}
  \centering
    \subfloat{\includegraphics[width=0.7\columnwidth]{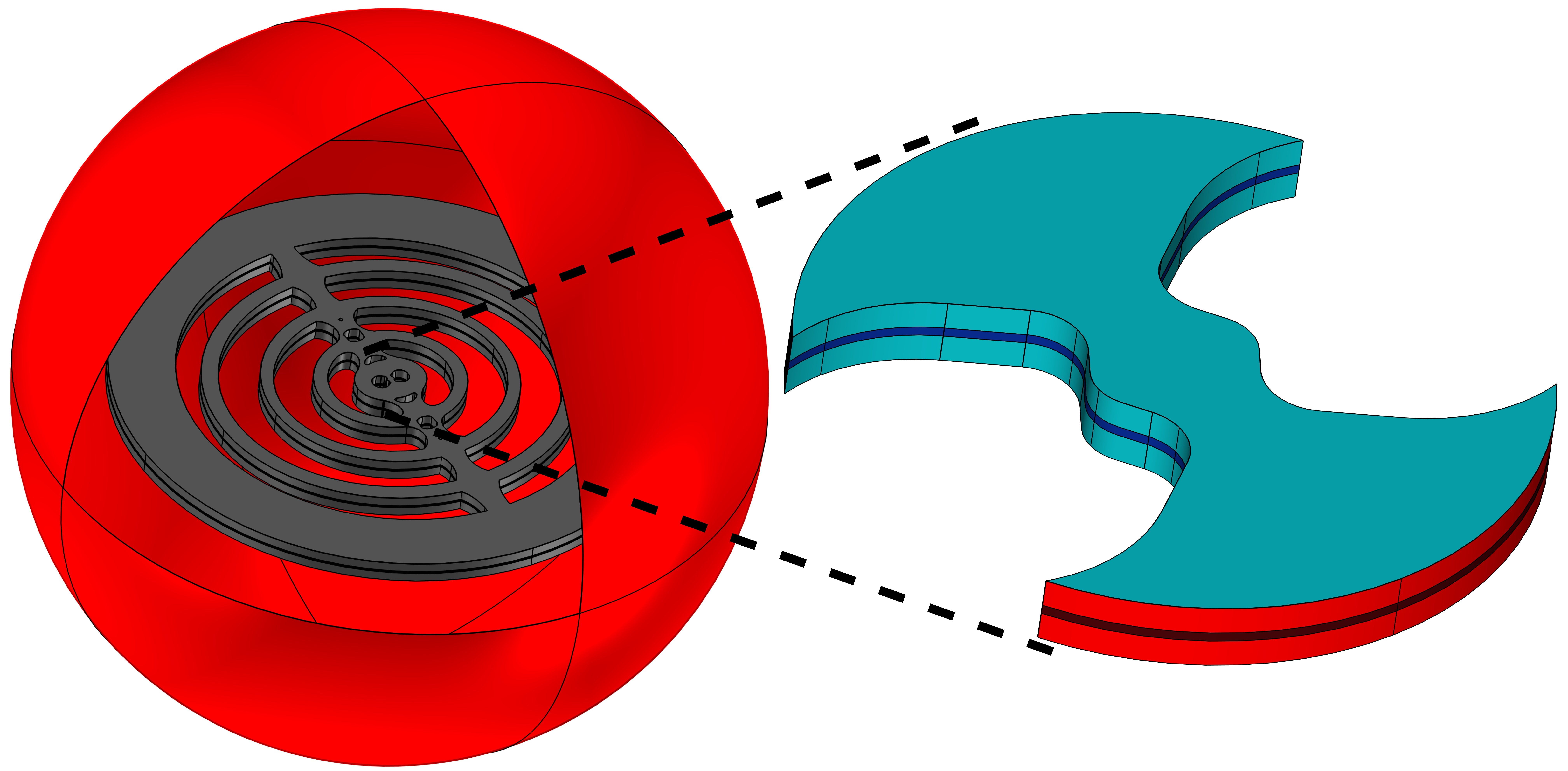}}
  \caption{Domains used for the calculation of photonic (left) and electronic (right) quasinormal modes. The right figure shows the well and barrier layers in a radius of $200 \, \mathrm{nm}$. Color coding indicates the boundary conditions. Red boundaries: radiation conditions, cyan boundaries: Dirichlet conditions.}
  \label{fig3}
\end{figure}

For the electronic QNMs $\tilde{\psi}_\nu$, with discrete index $\nu$, we solve the effective mass Schr\"{o}dinger equation for electrons and holes and impose the scalar Sommerfeld radiation condition in the plane of the quantum well,
\begin{equation}
-\frac{\hbar^2}{2}\nabla^2\left( \frac{1}{m_{\text{e}/\text{h}}^*}\tilde{\psi}_\nu(\mathbf{r}) \right) + V_{\text{e}/\text{h}}(\mathbf{r})\tilde{\psi}_\nu(\mathbf{r}) = \tilde{E}_\nu\tilde{\psi}_\nu(\mathbf{r}), \;\;\;\; \hat{\mathbf{r}}\cdot \nabla\tilde{\psi}_\nu(\mathbf{r}) \rightarrow  \text{i}\tilde{k}_\nu\tilde{\psi}_\nu(\mathbf{r}) , \; |\mathbf{r}|\rightarrow \infty,
\end{equation}

\noindent where $\hbar$ is the reduced Planck constant, $m_{\text{e}/\text{h}}^*$ is the electron (e) or hole (h) effective mass, $V_{\text{e}/\text{h}}(\mathbf{r})$ is the corresponding potential, and $\tilde{E}_\nu$ is the complex eigenenergy, for which the real and the imaginary parts correspond, respectively, to a characteristic resonance frequency $\omega_\nu = \rm{Re}(\tilde{E}_\nu)/\hbar$ and decay rate $\gamma_\nu = -\rm{Im}(\tilde{E}_\nu)/\hbar$.  The wave vector $\tilde{k}_\nu$ is related to the complex eigenenergy via the band diagram of the infinitely extended quantum well, and we apply the radiation condition to the in-plane edges of the domain as illustrated with red in Fig. \ref{fig3}. The electronic wavefunction vanishes at semiconductor-air interfaces, so we apply Dirichlet conditions to the remaining boundaries, shown in the same figure in cyan. In practical calculations, we use an analytical continuation of the band structure on the real energy axis \cite{Kristensen2014}. As with optical QNMs, the complex wave vector results in a divergence of the field away from the resonator, which, in turn, requires an analogous normalization. We further discuss the radiation conditions and their implementations, as well as the normalizations, in Appendix \ref{AppendixA}.

\subsection{System dynamics}

Based on the optical and electronic QNMs, we now calculate the strength of the light-matter interaction. To this end, we follow the approach of Franke \emph{et al.} \cite{Franke2019} to set up a model for the quantized electromagnetic field in terms of a single optical QNM with index $\mu=\text{c}$. In this way, the electric field operator can be written in terms of bosonic creation and annihilation operators $a_{\rm{c}}^\dagger$ and $a_{\rm{c}}$, respectively, as $\mathbf{E(r},t)=\mathrm{i}\sqrt{\frac{\hbar\omega_{\rm{c}}}{2\epsilon_0}}\mathbf{\tilde{f}}_{\rm{c}}(\mr )a_{\rm{c}}(t) + \text{h.a.}$, in which $\epsilon_0$ is the permittivity of free space, and where $a_{\rm{c}}^\dagger$ and $a_{\rm{c}}$ obey a Langevin-type equation with dissipation and a fluctuating noise term. In a similar manner, and in order to account for the leakage of the excitons from the central QD region, we treat the excitons in the single-excitation limit as harmonic oscillators with bosonic creation and annihilation operators $\sigma_{\rm{x}}^\dagger$ and $\sigma_{\rm{x}}$ and add appropriate terms to the equation of motion to capture the dissipation. We then write the dynamical equations of the system operators as
\begin{align}
\frac{d}{dt} a_{\rm{c}}(t) &= -\frac{\rm{i}}{\hbar}[a_{\rm{c}}(t),H] - \gamma_{\rm{c}} a_{\rm{c}}(t) + F_{\rm{c}} (t) \\
\frac{d}{dt} \sigma_{\rm{x}}(t) &= -\frac{\rm{i}}{\hbar}[\sigma_{\rm{x}}(t),H] - \gamma_{\rm{x}} \sigma_{\rm{x}}(t) + F_{\rm{x}} (t)
\end{align}

\noindent where $H = \hbar \omega_{\rm{c}} a_{\rm{c}}^\dagger(t)a_{\rm{c}}(t)  + \hbar \omega_{\rm{x}} \sigma_{\rm{x}}^\dagger (t)\sigma_{\rm{x}} (t)+ H_\mathrm{int}$, in which $H_\mathrm{int}$ is the interaction Hamiltonian and $\tilde{\omega}_{\rm{x}} = \omega_{\rm{x}} -\rm{i}\gamma_{\rm{x}}$ is the complex eigenfrequency of the exciton. The operators $F_{\rm{c}}$ and $F_{\rm{x}}$ are fluctuating source terms which act to preserve the commutation relations $[a_{\rm{c}}(t), a_{\rm{c}}^\dagger(t)]=[\sigma_{\rm{x}}(t),\sigma_{\rm{x}}^\dagger(t)]=1$ in the presence of loss \cite{Gardiner04, Franke2019}. For the excitons, we consider the lowest energy electron and hole states, which we label with indices $\nu=\text{e}$ and $\nu=\text{h}$ to ease the notation, and we focus on the strong confinement regime, in which the confinement energy is much larger than the binding energy due to Coulomb interaction. The complex exciton frequency can then be written as the sum of the complex electron and hole frequencies and the real frequency corresponding to the bandgap, $\omega_\text{x} = \omega_\text{e}+\omega_\text{h} + E_\text{BG}/\hbar, \gamma_\text{x}=\gamma_{\rm{e}} + \gamma_{\rm{h}}$, and the exciton wavefunction with units of inverse volume can be found as the product of the normalized electron and hole wavefunctions, $\chi(\mr)=\tilde{\psi}_\text{e}(\mr)\tilde{\psi}_\text{h}(\mr)$. The exciton wave function is associated with a polarization density $\mathbf{P}$, which we now assume can be expanded on a single excitonic wavefunction as $\mathbf{P(r,}t) = \mathbf{d} \, \chi (\mathbf{r})\sigma_{\rm{x}}^\dagger(t) + \text{h.a.}$, where $\mathbf{d}=d\,\mathbf{\hat{e}}_\text{d}$ is the dipole moment of the quantum well material with magnitude $d$. For the light-matter interaction, we can then sum all spatial contributions by integrating the quantity $\mathbf{P(r},t)\cdot \mathbf{E(r},t)$ and write the interaction Hamiltonian as
\begin{equation}
H_\mathrm{int}\mathbf{(r},t) = \int \mathbf{P(r},t) \cdot \mathbf{E(r,}t) \, \text{d}V.
\end{equation}
\noindent Working in the single mode limit for both the electromagnetic field and the exciton state, and applying the rotating wave approximation keeping only the excitation conserving terms, we get the familiar form $H_{\rm{int}}= \hbar g a_{\rm{c}}^\dagger \sigma_{\rm{x}} + \text{h.a.}$, in which the coupling strength is given as
\begin{equation}
\hbar g = \text{i} d \sqrt{\frac{\hbar \omega_{\rm{c}} }{2\epsilon_0}} \;  \int \chi(\mr) \; \mathbf{\hat{e}}_\text{d} \cdot \mft_\text{c}(\mr) \text{d}\mathbf{r}.
\label{Eq:hbar_g}
\end{equation}
We note that in the dipole limit $\chi(\mr) = \delta(\mr-\mr_\text{d})$, where $\mr_\text{d}$ is the dipole position, we recover the known dipole coupling strength as in Ref.~\cite{Franke2019}. Owing to the radiation condition and the resulting divergent nature of the QNMs, the integral in Eq.~(\ref{Eq:hbar_g}) formally diverges as a function of integration domain size. The integral can, in principle, be regularized by use of complex coordinate transforms, for example~\cite{Kristensen2015}, but for the calculations in this work, the quality factors are sufficiently large that such a procedure in practice is not needed; see Appendix~\ref{AppendixA} for details. The bulk dipole moment can be expressed in terms of the momentum matrix element $p_\text{cv}$ as $d = e |p_{\mathrm{cv}}|/m_0 \, \omega_\text{x},$ where $e$ is the electronic charge, $m_0$ is the electron rest mass and $|p_\mathrm{cv}|^2= m_0 E_p(x)/2$, where $E_p(x)$ is the Kane energy~\cite{Stobbe2012}. For a typical value of $E_p = 20 \, \mathrm{eV}$ in common quantum well semiconductors~\cite{Vurgaftman2001}, we find $p\sim 1.1 \, \mathrm{e \; nm}$, which is similar to measured values in self-organized QDs~\cite{Eliseev2000, Silverman2003}, and close to the value used in Ref.~\cite{Cartar2017}. In the following analysis, we therefore use this value as a representative size of the dipole moment. As noted above, we focus on transitions involving the electron and hole states with the lowest energies, which are expected to contribute most to radiative transitions. For a discussion on other states and selection rules, see Appendix \ref{AppendixB}.

\subsection{Radiative rate and efficiency}

To assess the usefulness of the proposed design, we now turn to modeling the expected light emission properties. Because of the proposed etching through the active material, we expect non-radiative decay channels to play an important role, and the radiative rate alone is likely an incomplete description of performance. For a more representative picture in the presence of additional decay channels, therefore, we consider the enhanced radiative rate in conjunction with the radiative quantum efficiency. To study the dynamics of the system and in order to quantify these figures of merit, we use the Lindblad formalism~\cite{Breuer2007, Rivas2012} and consider the density matrix of the interacting cavity photon-exciton system in the single-excitation subspace spanned by the Fock states corresponding to either one photon or one exciton. The density matrix $\rho(t)$ evolves according to the equation
\begin{equation}
\frac{d\rho}{dt} = -\frac{\text{i}}{\hbar}[H,\rho] + \Gamma_\text{c} D(a_\text{c},\rho) + \Gamma_\text{nr} D(\sigma_\text{x},\rho) + 2 \gamma^* D(\sigma_\text{x}^\dagger \sigma_\text{x},\rho) \; ,
\label{eq9}
\end{equation}
in which $D(O,\rho) = O\rho O^\dagger - \frac{1}{2}\left(O^\dagger O \rho + \rho O^\dagger O \right)$ denotes the Lindblad dissipator for a generic operator $O$, and $\Gamma_\text{c}=2\gamma_{\rm{c}}$. For the exciton, the total non-radiative population decay rate $\Gamma_\text{nr} = 2\gamma_\text{x} + \Gamma_\text{s}$ is the sum of the electron and hole QNM tunneling rates and a non-radiative rate of surface population recombination. In addition, we consider the coherences of the excitons to be subject to pure dephasing with a characteristic rate $\gamma^*$, which we model via the operator $\sigma_\text{x}^\dagger \sigma_\text{x}$~\cite{BundgaardNielsen2021}. While the cavity decay rate and the tunneling rates of the electrons and holes are found from the calculated QNMs, $\gamma^*$ and $\Gamma_\text{s}$ are taken as phenomenological parameters in this model. Depending on the relative size of these rates, we find different values for the radiative rate, enhancement, and radiative quantum efficiency. Assuming an initial state with one exciton and zero photons in the bad cavity and the weak coupling limit, and by applying the adiabatic approximation, the total emitter decay rate for zero detuning between the electronic transition frequency and the cavity resonance frequency is~\cite{Mrk2015}
\begin{equation}
\Gamma_{\rm{tot}} \approx \Gamma_\text{nr} + \Gamma_\text{rad},
\label{eq10}
\end{equation}
where $\Gamma_\text{nr} = \Gamma_{\rm{tx}} + \Gamma_{\rm{s}}$, and 
\begin{equation}
\Gamma_\text{rad}= \frac{4g^2}{\Gamma_\text{c} + \Gamma_\text{nr} + 2\gamma^*}
\label{Eq:Gamma_rad}
\end{equation}
is the radiative rate. In this limit, the radiative quantum efficiency is typically defined as the ratio of the radiative to the total decay rate \cite{Hughes2019},
\begin{equation}
\eta = \frac{\Gamma_{\rm{rad}}}{\Gamma_\text{nr} + \Gamma_{\rm{rad}}}.
\label{eq11}
\end{equation}

\noindent We present this in combination with the enhanced radiative rate as common and important metrics to fairly judge the design performance in the presence of realistic non-radiative rates introduced by the etching process. However, we point out that the overall efficiency of the implemented device will, in practice, also depend on other factors, such as the collection efficiency to far-field optics or coupled waveguides \cite{Cui2005, Bulgarini2012, Matsuzaki2021,  Jacobsen_Niels_Beta2023}.

\subsection{Varying the design}
\label{sec3_4}
\begin{figure}[b]
  \centering
    \subfloat{\includegraphics[width=0.75\columnwidth]{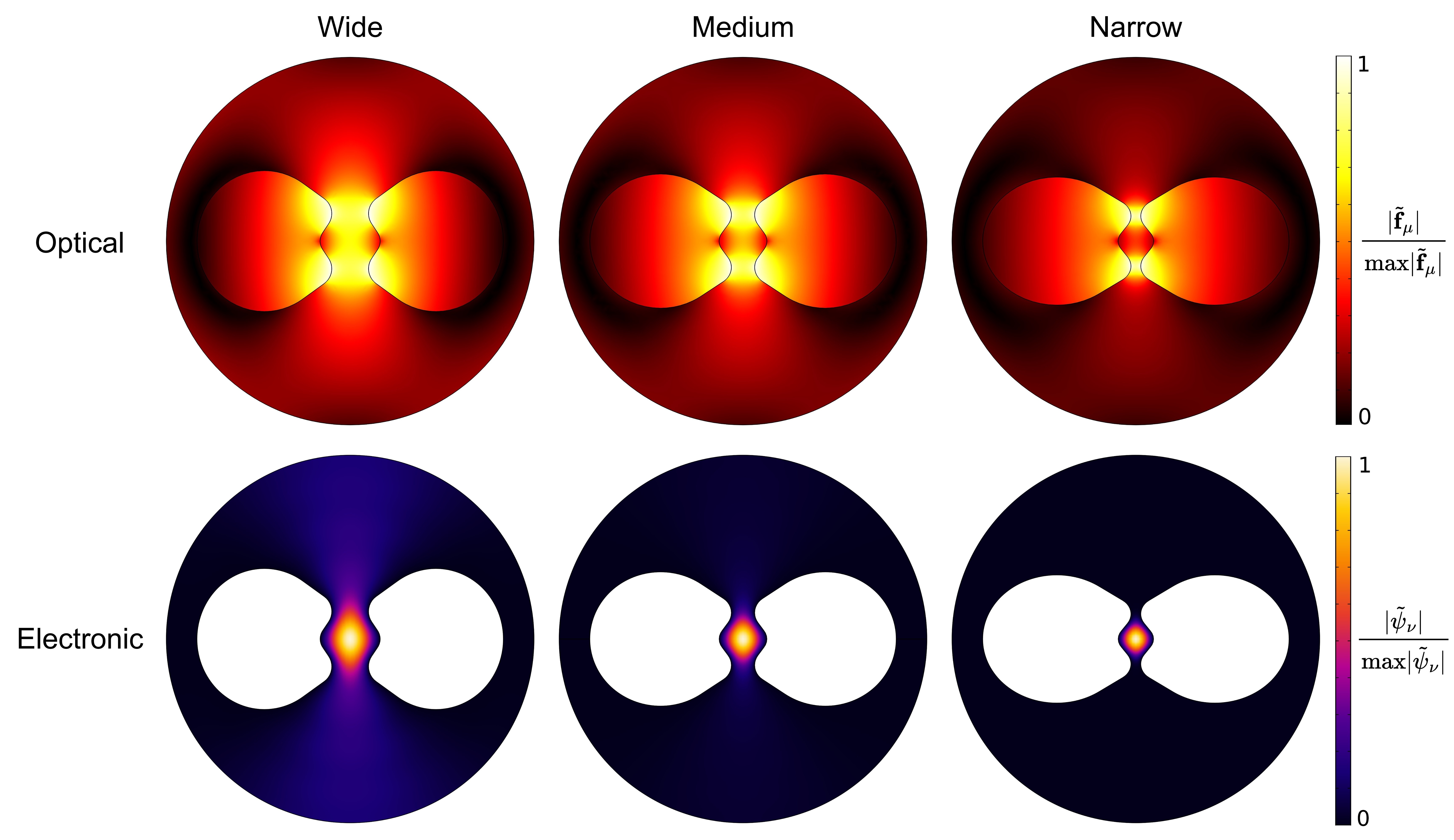}}
  \caption{Variations of the central part of the LDQD geometry to gauge the effect on the radiative rate of the transition. Shown are the relative norm of the electric field of the optical field (upper row) and the relative norm of the electronic state (bottom row) for the different designs. From left to right, we term the three designs: wide, medium, and narrow, with corresponding $w/d$ ratios $61.2\%, 48.7\%$, and $27.6\%$.}
  \label{fig4}
\end{figure}

In this section, we consider variations of the geometry to assess how the coupling strength and the lifetimes change, and to get some insight into potential optimization strategies. We consider three different designs of the LDQD with varying bridge/width ratios $w/d = \{ 61.2\%, 48.7\%, 27.6\% \}$, as shown in Fig. \ref{fig4}; see Fig. \ref{fig2}(b) for the definitions of $w,d$. The material parameters for this study can be found in Appendix \ref{AppendixA}.

Table \ref{Table1} summarizes the calculated tunneling rates of the electron and hole states, the cavity decay rate, and the coupling strength for the three designs. We find a clear trade-off between the quality factor of the photonic and the electronic QNMs. As the ratios $w/d$ decrease, the radiative loss of the electronic states decreases as well, whereas the optical cavity shows the opposite trend.

\begin{table}[!htbp]
\centering
\begin{tabular}{c | D{,}{\pm}{-1} D{,}{\pm}{-1} D{,}{\pm}{4.4}}
\multicolumn{1}{c|}{} & \multicolumn{3}{c}{\textbf{Calculated rates ($\rm{\mu eV}$)}} \\ \hline
{Rate} & \multicolumn{1}{c}{$\;\;\;\;\;\;$ Wide $\;\;\;\;\;\;$} & \multicolumn{1}{c}{$\;\;\;\;\;$ Medium $\;\;\;\;\;$} & \multicolumn{1}{c}{$\;\;\;\;\;\;$ Narrow $\;\;\;\;\;\;$} \\
\hline
$2\hbar \gamma_\text{e}$  & 178 , 18 & 40.8 , 1.8  &  (38 , 2) \cdot 10^{-3}  \\
$2\hbar \gamma_\text{h}$  & 26, 10 &  5.2 , 0.8  &  (4.8 , 0.4) \cdot 10^{-3} \\
$\hbar \Gamma_\text{c}$              & 163.9 , 0.8 & 587 , 74   & 1547 , 8   \\
$\hbar g$                   & 93.5 , 13.2 & 92.4 , 0.22 & 78.87 , 0.88   \\
\end{tabular}
\caption{Computed rates for the designs shown in Fig. \ref{fig4}. Tabulated are the tunneling rates of the electron and hole population $2 \gamma_\text{e/h}$, the cavity population decay rate $\Gamma_\text{c}$, and the calculated coupling strength $g$ for the three designs, cf. Fig.~\ref{fig4}. All quantities are in units of $\rm{\mu eV}$, and the assumed dipole moment of the state is $d=1.1 \, \rm{e \; nm}$.}
\label{Table1}
\vspace{-2mm}
\end{table}

Based on the data in Table \ref{Table1}, we can estimate the radiative rate and the efficiency of the device by assuming additional decay rates as introduced, for example, by the etching process. Table \ref{Table2} shows four representative cases for these rates. The first row shows the radiative rate in the absence of all non-radiative rates and dephasing, including the tunneling rates, $\Gamma_\text{tx} + \gamma^* + \Gamma_\text{s} = 0$, to illustrate the maximum possible radiative rates if even the tunneling rates were to be eliminated. These values are then only limited by $g$ and the cavity Q-factor and correspond to the Purcell limit, similar to Eq.~(\ref{eq1}). The second row includes the tunneling rates but assumes negligible surface recombination and pure dephasing, which may be achieved by proper surface passivation. Then, the only decay in the system is radiative decay and the tunneling of the electronic states, $2\gamma_\text{x} + \Gamma_\text{rad}$. This corresponds to a practical upper bound to the possible efficiency. This total decay rate corresponds to lifetimes on the order of $2.2 \, \rm{ps}$, $6.6 \, \rm{ps}$, and $40.8 \, \rm{ps}$ for the three designs.

\smallskip

In the next two rows, we show the possible extremes for a total added rate of $\hbar \Gamma_{\rm{s}} + \hbar \gamma^* = 100 \rm{\mu eV}$: either a pure non-radiative recombination rate or a pure dephasing rate. The chosen value of $100 \rm{\mu eV}$ is representative of cryogenic operation, based on the measured linewidth of etched QDs \cite{Verma_AIP_2011}. Whereas non-radiative recombination and pure dephasing contribute similarly to broadening the emission spectrum and have the same effect on the radiative rate according to Eq.~(\ref{Eq:Gamma_rad}), they have a different impact on the efficiency, as seen from Eq. (\ref{eq11}), as well as in the results of Table \ref{Table2}. In addition to this observation, while the coupling strengths remain similar, the optical decay rate
changes drastically between the designs, greatly influencing the achievable enhanced radiative rate, cf. Eq.~(\ref{Eq:Gamma_rad}). This combination of factors further accentuates the trade-off between the decay rates of the optical and the electronic QNMs and its relevance in design. In particular, while the limited diffusion away from the QD in the narrow design dramatically increases the possible efficiency, it still results in a smaller radiative rate, and therefore a larger sensitivity to surface recombination. With a successful passivation strategy, it will be possible to limit surface recombination, thereby approaching the upper bounds of efficiency and radiative rate for a specific design.

\begin{table}[!htbp]
\centering
\begin{tabular}{ c | c c c | c c c }
\multicolumn{1}{c|}{} & \multicolumn{3}{c|}{\textbf{$\eta$ ($\%$})} & \multicolumn{3}{c}{\textbf{$\hbar\Gamma_\text{rad}$ ($\rm{\mu eV}$})} \\ \hline
$\;\;\;$ Additional decay rates $\;\;\;$ & $\;\;$ Wide $\;\;$  & $\;\;$ Medium $\;\;$ & $\;\;$ Narrow $\;\;$ &  $\;\;$ Wide $\;\;$   & $\;\;$ Medium $\;\;$ & $\;\;$ Narrow $\;\;$ \\ \hline
$\Gamma_\text{tx} + \gamma^* + \Gamma_\text{s} = 0 \;\;\;\;\;\;\;$ & $100$	 & $100$  & $100$ & $213$ & $58$ & $16$ \\
$\;\;\;\;\;\;\;\;\;  \gamma^* + \Gamma_\text{s} = 0 \;\;\;\;\;\;\;$         & $32$	 & $54$  & $99.7$ & $95$ & $54$ & $16$ \\
$\hbar\Gamma_{\rm{s}} = 100 \, \mu \text{eV}$   & $20$  & $24$  & $13$ & $75$ & $47$ & $15$   \\
$\hbar\gamma^*= 100 \, \mu \text{eV}$           & $23$	 & $47$  & $99.7$ & $62$ & $41$ & $14$ \\
\end{tabular}
\caption{Efficiency $\eta$ ($\%$), as defined in Eq. (\ref{eq11}), and enhanced radiative rate $\hbar\Gamma_\text{rad}$ $(\rm{\mu eV})$, as in Eq. (\ref{Eq:Gamma_rad}), for the considered variations $w/d$ of the design, cf. Fig.~\ref{fig4}. Both metrics are shown assuming three different cases of surface recombination rate $\Gamma_{\rm{s}}$ and pure dephasing rate $\gamma^*$.}
\label{Table2}
\end{table}

\hspace{1mm}

\vspace{-5mm}

\section{Conclusion}
\label{Sec:Conclusion}

We have proposed a new approach for the deterministic fabrication of a QD in a region of deep sub-wavelength confinement of light by modifying a dielectric bowtie cavity with an embedded quantum well to simultaneously confine electron and hole states in the region of the optical hot spot. We quantified the radiative enhancement in the bad cavity limit and found that the approach is promising for overcoming the high non-radiative rates in etched QDs, thus achieving deterministic fabrication of a functional, single QD in an optical cavity. With the development of more precise lithography and better passivation techniques, this may provide a competitive method to deterministically fabricate optical cavities with embedded single QDs for integrated photonics applications featuring both high radiative enhancement and efficiency.

\bigskip

\section*{Funding}
This work was supported by the Danish National Research Foundation through NanoPhoton - Center for Nanophotonics, grant number DNRF147. Emil Vosmar Denning acknowledges support from Independent Research Fund Denmark through an International Postdoc Fellowship (Grant No. 0164-00014B).

\bibliographystyle{ieeetr}

\bibliography{Bibliography}

\newpage

\appendix

\section{Calculations}
\label{AppendixA}

\subsection{Set-up, radiation conditions and normalizations}

\begin{figure}[b]
  \centering
    \subfloat{\includegraphics[width=0.85\columnwidth]{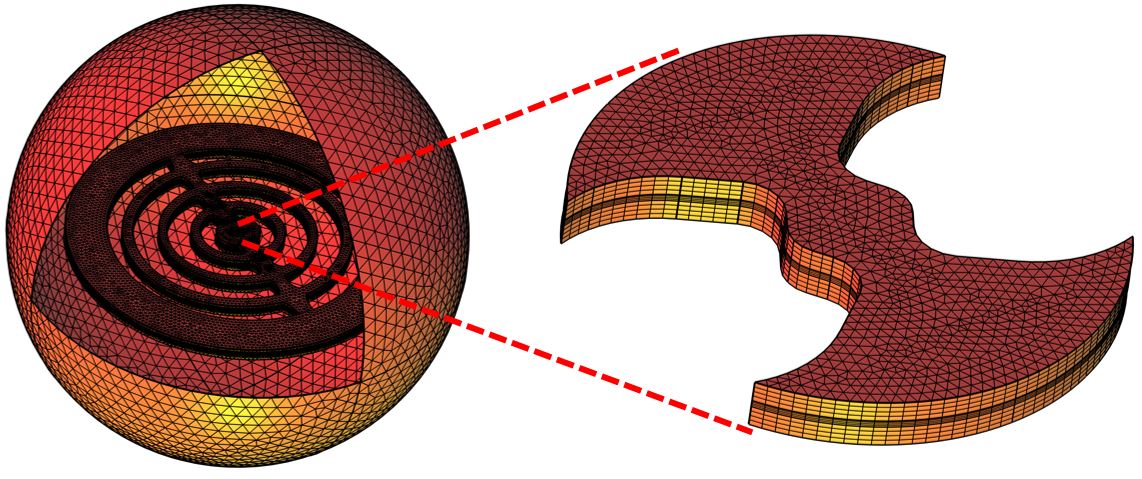}}
  \caption{Finite element discretization for the electromagnetic (left) and the electronic (right) problem. We use tetrahedral elements for the former, while for the latter, we use triangular prismatic elements.}
  \label{figA1}
\end{figure}

We use different meshes for the optical and electronic calculations, as illustrated in Fig.~\ref{figA1}. For the electromagnetic problem, we consider the entire free-standing cavity surrounded by a spherical air domain with a numerical radiation condition on its surface. For the electrons and holes, we limit the domain to a region around the bowtie and apply either radiation conditions or Dirichlet boundary conditions as illustrated in Fig.~\ref{fig3} of the main text. For the optical QNMs, we implement the radiation condition as a boundary condition on the edge of the domain with normal vector $\hat{\mathbf{n}}$ as
\begin{align}
\hat{\mathbf{n}}\times\left[\nabla\times \mft_\mu(\mr)\right] +{\text{i}} n_{\text{B}} \tilde{k}_\mu \hat{\mathbf{n}} \times \left[ \mft_\mu(\mr)  \times \hat{\mathbf{n}} \right] = 0.
\end{align}
For the electronic states, we assume that the solutions far from the LDQD approach the propagating states in an infinitely extended quantum well between two barrier layers. In this way, we define a radiation condition for the scalar problem of the form
\begin{equation}
\hat{\mathbf{r}}\cdot \nabla\tilde{\psi}_\nu(\mathbf{r})-  \text{i}\tilde{k}_\nu\tilde{\psi}_\nu(\mathbf{r})=0,
\end{equation}
where the complex wave vector is found by analytic continuation of the inverted band structure $\tilde{k}_\nu(\tilde{E}_\nu)$ for the infinite quantum well~\cite{Kristensen2014}. For the fundamental state in the LDQD, which is even in $z$, we assume that the relevant band is the fundamental even state in the well. In practice, we approximate this band by a quadratic polynomial $k(E)$ close to the energy of interest and approximate the boundary condition using this polynomial.

\smallskip

\noindent In practice, we normalize the optical QNMs using the condition~\cite{Muljarov_OL_43_1978_2018, Kristensen2020}
\begin{align}
\frac{1}{2\epsilon_0}\int_V\epsilon_0\epsilon_\text{r}(\mr)\mft_\mu(\mr)\cdot\mft_\mu(\mr) - \mu_0 \mgt_\mu(\mr)\cdot\mgt_\mu(\mr)\,\ud V +\frac{\text{i}}{2\epsilon_0\tlo_\mu}\int_{\partial V} &\left[\mft_\mu(\mr)\times[\mr\cdot\nabla\mgt_\mu(\mr)]\right.\nonumber\\ 
&\left.- [\mr\cdot\nabla\mft_\mu(\mr)]\times\mgt_\mu(\mr)\right]\cdot\mathbf{\hat n}\; \ud A=1,
\end{align}
where $\mu_0$ is the permeability of free space, and $\mgt_\mu(\mr)=\nabla\times\mft_\mu(\mr)/[\text{i}\tlo_\mu\mu_0]$ denotes the magnetic field QNM of index $\mu$. The volume $V$ is taken as the entire calculation domain, and $\partial V$ is the outermost surface. Owing to the relatively large quality factor, the contribution from the surface integral is negligible. For the scalar electronic states, we use the condition
\begin{equation}
\int_V\tilde{\psi}_\nu(\mr) \tilde{\psi}_\nu(\mr) \;  dV + \frac{\rm{i}}{2\tilde{k}_\nu} \int_{\partial V} \tilde{\psi}_\nu(\mr) \tilde{\psi}_\nu(\mr) \; \ud A=1,
\end{equation}
where the volume $V$ is the calculation domain for the electronic problem, as illustrated in Fig.~\ref{figA1}. The surface integral is taken over the entire boundary of $V$, but in practice has support only on two of the edges as illustrated in Fig.~\ref{fig3}. In these directions, the fields diverge as a function of distance, and the surface integral can be viewed as a convenient regularization derived from a far-field approximation as in Ref.~\cite{Lai1990}. This is more than adequate given these states' relatively large quality factors, for which the surface integral contribution is negligible at all practical calculation domain sizes. We note, however, that the divergence can in principle be handled perfectly by the elegant approach of Ref.~\cite{Muljarov2010} or by a suitable additional regularization as discussed in Ref.~\cite{Kristensen2015}.

\subsection{Convergence studies}

\begin{figure}
  \centering
    \subfloat{\includegraphics[width=0.95\columnwidth]{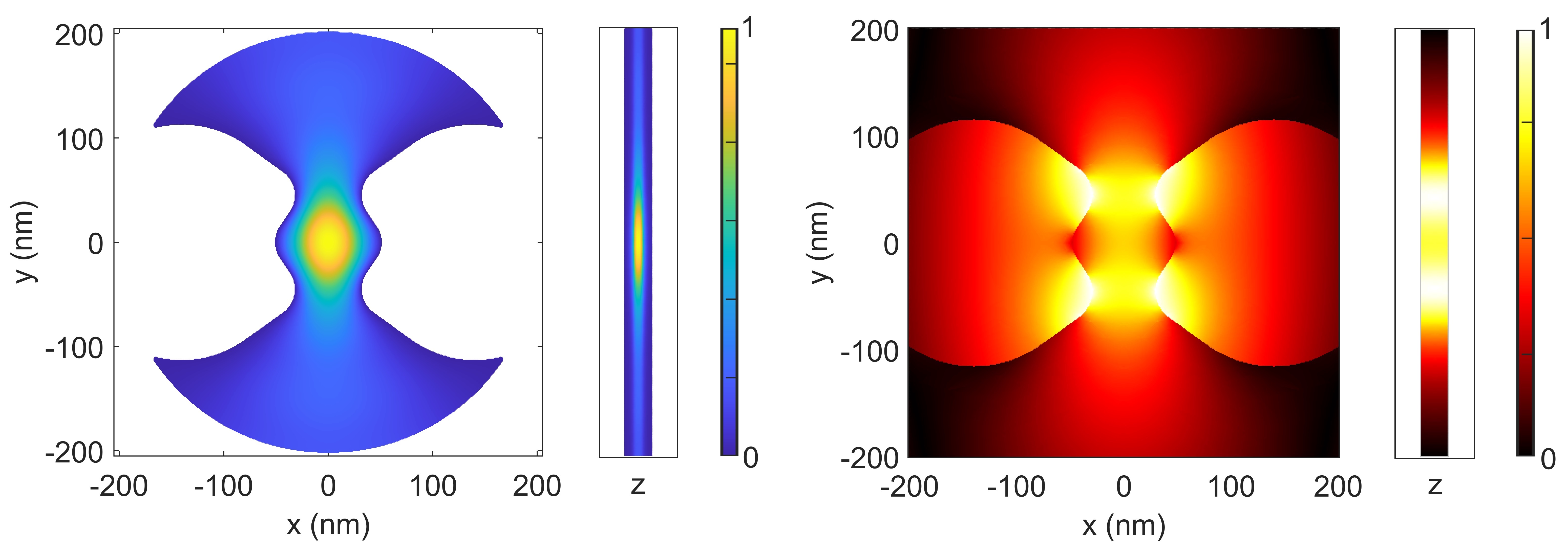}}
  \caption{Cartesian extraction grid used for the overlap integral. The left panels show the absolute value of the electronic state in the $xy$ and $yz$ planes in the domain shown in Fig. \ref{fig3}. The thickness in the $z$-direction is $38.3$\,nm. The right panels show the $y$-component of the optical field in the same region.}
  \label{figA2}
\end{figure}

Two important parameters are involved when considering the convergence of quantities associated with the individual QNM calculations: the calculation domain size and the discretization. The distance to the computational domain boundaries affects the accuracy of the imposed numerical radiation conditions and becomes more accurate for larger domains. At the same time, the discretization also affects the computational result since it controls the numerical approximation of the continuous problem in terms of local basis function. For the convergence studies, we consider a systematic variation of these parameters. To this end, we choose an initial coarse mesh and small domain size and gradually increase the domain size, defined by the radius of the surrounding spherical domain in the optical case, or by the radius of the circle surrounding the LDQD in the electronic calculations, as shown in Fig. \ref{figA1}. We then repeat these calculations by successively refining the previous meshes while keeping the order of the basis functions fixed. We use a different approach for the optical and electronic calculations. For the optical QNMs, we use isoparametric and quadratic tetrahedral elements, and each refinement of the mesh results by splitting the elements along their longest side, roughly doubling the total number of elements in the domain. For the electronic states, we use triangular prismatic elements, as illustrated in Fig. \ref{figA1}, with linear approximations to the field and quadratic approximations to the geometry. To refine this mesh, we scale the number of elements by a factor of $\sqrt{2}$ along x and y, and double the number of elements along the vertical direction. This results in a refined mesh with approximately four times the elements of the previous iteration.

For the calculation of $g$, which involves an overlap integral of the different QNMs, we pair up results from each mesh refinement. For the coarsest mesh, for example, we use a set of 10 domain sizes for both electronic and optical QNMs, and calculate the overlap integral for the corresponding domain sizes. We use domains with radii of $200-300 \, \rm{nm}$ in steps of $10 \, \rm{nm}$ for the electronic states, and $(2.5-3.5)\lambda_0$ in steps of $0.1 \lambda_0$, where $\lambda_0 = 1550 \, \rm{nm}$, for the optical fields. To calculate $g$, we then extract the QNMs on a cartesian grid and use a simple quadrature to calculate the overlap integral. Fig. \ref{figA2} shows the relative norms of the extracted electron state and of the optical field along $y$ for a square region $400 \, \rm{nm} \times 400 \, \rm{nm}\times 38.3 \, \rm{nm}$. For the convergence studies, we estimate the errors in the same way as described in the supplementary information of \cite{Kountouris_22}. For $g$ we could also include an estimated error associated with the discretization of the extraction grid, for which we used a discretization of $1 \,  \rm{nm}$ in $x$ and $y$ and $0.25 \, \rm{nm}$ in $z$. The effect of the extraction grid, however, was found to be small compared to other numerical errors. We note that despite the QNMs being spatially divergent, they have sufficiently high quality factors that the effect is not visible even for a domain spanning $2 \, \rm{\mu m}$. Therefore, we consider any residual error stemming from the QNM divergences to be negligible in these calculations.

\bigskip

\subsection{Material parameters}

For the optical modeling, we assumed a dielectric membrane made of indium phosphide with refractive index 3.16 at room temperature and a total thickness of $240 \, \rm{nm}$, including the well structure. To simplify the numerical calculations, we have neglected the difference in the optical properties of the quantum well structure and used a constant refractive index throughout the material. For the electronic calculations, we assumed $16.4 \, \rm{nm}$ thick barriers made of aluminum indium gallium arsenide (AlInGaAs) on both sides of a $5.5 \, \rm{nm}$ thick gallium indium phosphide (GaInAsP) quantum well with the parameters shown in table \ref{Table4}.

\begin{table}[!htbp]
\centering
\begin{tabular}{ c c c  }
\multicolumn{3}{c}{\textbf{Parameters for the effective mass Schr\"{o}dinger equation}} \\ \hline
{ } & \multicolumn{1}{c}{$\;\;\;\;\;\;\;\;\;\;\;\;\;\;\;\;\;$ Barrier $\;\;\;\;\;\;\;\;\;\;\;\;\;\;\;\;\;$} & \multicolumn{1}{c}{$\;\;\;\;\;\;\;\;\;\;\;\;\;\;\;\;\;$ Well $\;\;\;\;\;\;\;\;\;\;\;\;\;\;\;\;\;$} \\
\hline
Material  & Al(x)In(y)Ga(1-x-y)As   & Ga(x)In(1-x)As(y)P(1-y) \\
Assumed x & 0.18   & 0.23 \\
Assumed y & 0.452  & 0.8625 \\
Thickness (nm) & 16.4  & 5.5 \\
Effective mass e & 0.0575 ${m_0}$ & 0.0394 ${m_0}$ \\
Effective mass h & 0.07 ${m_0}$ & 0.3705 ${m_0}$ \\
Bandgap (eV) & 0.7987 & 0.7987 \\
Potential barrier e (eV) & 0.25  & {-} \\
Potential barrier h (eV) & 0.125 & {-} \\
\end{tabular}
\caption{Parameters used for the quantum well and barrier materials in the effective mass Schr\"{o}dinger equation.  Here, $m_0$ denotes the electron rest mass.}
\label{Table4}
\end{table}

\newpage

\section{Additional electronic states and symmetry considerations}
\label{AppendixB}

\begin{figure}[b]
  \centering
    \subfloat{\includegraphics[width=0.7\columnwidth]{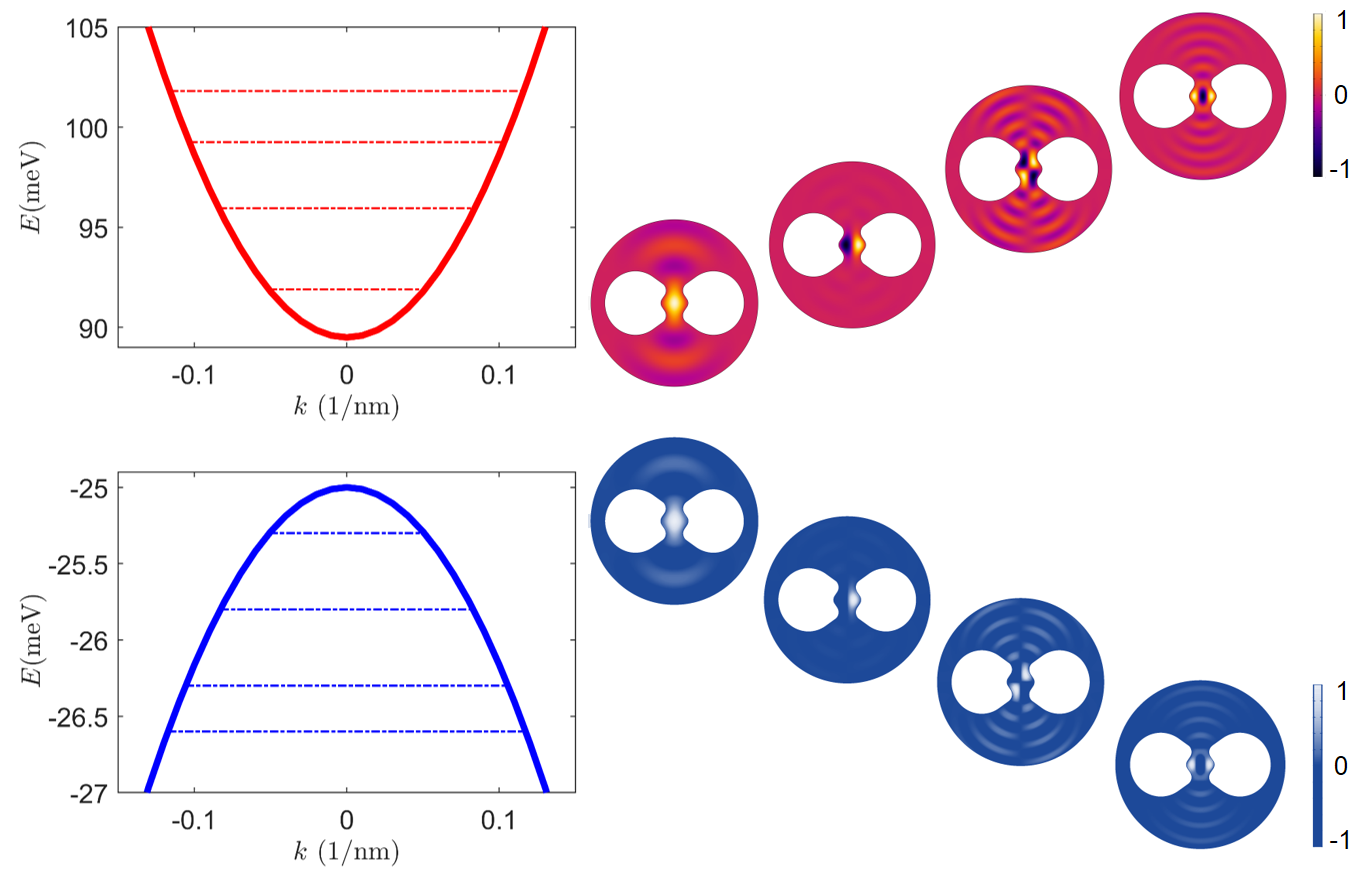}}
  \caption{Different electronic states supported in the QD geometry and their energies relative to the fundamental bands in the infinitely extended quantum well. The left panels show the fundamental electron (red) and hole (blue) bands of the extended well, along with horizontal lines indicating the real part of the energy of the first four states that are even with respect to the plane in the center of the quantum well. The figures on the right show the relative real part of the associated states.}
  \label{figA3}
\end{figure}

The geometry supports a number of discrete electronic states, with energies and energy separations that primarily depend on the QD width. For the wide design, the energies and field profiles of the first four states with even symmetry relative to the plane in the middle of the structure are represented in Fig.~\ref{figA3}. For this design, the energy separation is approximately $3-4 \, \rm{meV}$ for electrons and $0.5 \, \rm{meV}$ for holes. Furthermore, the fundamental state transition lies approximately $2.7 \, \rm{meV}$ higher than the lowest bands transition in the infinitely extended quantum well. For the other two designs, the picture is similar, but the energies of individual states are higher, and the energy separations between the different states are larger (not shown).

\smallskip

For the analysis, we have considered the recombination of excitons formed by the lowest-lying electron state and the highest-lying hole state, but one may ask if other combinations of states are possible. Indeed, multiple combinations are possible, as can be seen from the overlap integral in the complex coupling strength, $g\propto \int \tilde{\psi}_\text{e}(\mr)\,\tilde{\psi}_\text{h}(\mr)\,\mathbf{\hat{e}}_\text{d} \cdot \mft_\text{c}(\mr)\, \ud V$. Based on this expression, we can find simple selection rules based on the symmetries of the electronic states and the vectorial component of the photonic field, since the product of the parity of the individual factors needs to be even in all three axes for the overlap integral to be non-zero. The $y$-component of the optical field is even in all three directions, but the $x$-component is odd in $x$ and $y$. For $\mathbf{\hat{e}}_\text{d}=\hat{y}$, the electron and hole states therefore need to both be of the same parity in all three directions, whereas for $\mathbf{\hat{e}}_\text{d}=\hat{x}$ they need to be of opposite parity in $x$ and $y$, but the same parity in $z$. 

\medskip

To assess whether we can excite specific transitions, we can compare the energies of the states with their corresponding linewidths. For the wide design, the linewidths of the electron and hole states can vary greatly, with a maximum FWHM linewidth of about $0.4 \, \rm{meV}$, while the cavity FWHM linewidth is $0.16 \, \rm{meV}$, meaning it might be possible to address individual transitions, assuming additional rates are small enough. For the other two designs, the energy separations are larger and the tunneling rates are much smaller, and the corresponding cavity linewidth, at $0.6$ meV and $1.8$ meV for each design, is still much narrower than each individual separation. The main limitation for addressing specific transitions is then the presence of non-radiative recombination and dephasing in realistic devices, which will broaden the excitonic linewidths.

\end{document}